\documentclass[preprint]{aastex}
\pdfoutput=1
\usepackage{natbib}
\usepackage{color}
\usepackage{aas_macros}
\font\manual=manfnt at 7pt \def\dbend{\hbox{\raise0.9ex\hbox{\manual\char127\hspace{0.6em}}}}

\providecommand{\e}[1]{\ensuremath{\times 10^{#1}}}

\newcounter{INTERNALionstage}

\def\gtsim{\mathrel{\hbox{\rlap{\hbox{\lower4pt\hbox{$\sim$}}}\hbox{$>$}}}}
\def\lesssim{\mathrel{\hbox{\rlap{\hbox{\lower4pt\hbox{$\sim$}}}\hbox{$<$}}}}

\def\erg{{\rm\thinspace erg}}

\DeclareMathAlphabet{\vib}{OML}{cmm}{m}{it}
\begin{document}

\title{Bootstrapping dielectronic recombination from second-row elements and the Orion Nebula}

\author{
N.~R. Badnell\altaffilmark{1}, 
G.~J. Ferland\altaffilmark{2,3}, 
T.~W. Gorczyca\altaffilmark{4},
D. Nikoli\'c\altaffilmark{5},
G.~A. Wagle\altaffilmark{2}, 
}

\altaffiltext{1}{University of Strathclyde, Glasgow, G4 0NG, UK}

\altaffiltext{2}{University of Kentucky, Lexington, KY 40506, USA}

\altaffiltext{3}{The Queen's University of Belfast, Belfast, BT7 1NN, UK}

\altaffiltext{4}{Western Michigan University, Kalamazoo, MI 49008, USA}

\altaffiltext{5}{Jet Propulsion Laboratory, California Institute of Technology, Pasadena, CA 91109, USA}

\begin{abstract}
Dielectronic recombination (DR) is the dominant recombination process
for most heavy elements in photoionized clouds.
Accurate DR rates for a species can be predicted when the positions of autoionizing states are known.
Unfortunately such data are not available for most third and higher-row elements.
This introduces an uncertainty that is especially acute for photoionized clouds,
where the low temperatures mean that DR occurs energetically through very low-lying autoionizing states.  
This paper discusses S$^{2+} \rightarrow$ S$^+$ DR,
the process that is largely responsible for 
establishing the [S~III]/[S~II] ratio in nebulae.
We  derive an empirical rate coefficient using
a novel method for second-row ions, which do have accurate data.
Photoionization models are used to reproduce the [O~III] /  [O~II] / [O~I] / [Ne~III] intensity
ratios in central regions of the Orion Nebula.  O and Ne have accurate
atomic data and can be used to derive
an empirical  S$^{2+} \rightarrow$ S$^+$ DR rate coefficient at $\sim 10^{4}$~K.
We present new calculations of the DR rate coefficient for
S$^{2+} \rightarrow$ S$^+$ and quantify how uncertainties in the autoionizing level positions
affect it.
The empirical and theoretical results are combined and we derive a simple fit
to the resulting rate coefficient at all temperatures for incorporation into spectral
synthesis codes.
This method can be used to derive empirical DR rates for other ions, provided that
good observations of several stages of ionization of O and Ne are available.

\end{abstract}


%\today

\section{Introduction}

Although numerical simulations are 
the best way to understand the message in astronomical spectroscopic observations, 
they can be no better than the underlying atomic and molecular data.  
Theoretical predictions of cross sections and rate coefficients provide 
the vast majority of the data now used.  
There are, however, still significant gaps in the database.  
Given the amount of effort that has been expended in the past 
it is inevitable that today's outstanding needs are also the most challenging ones.  
Dielectronic recombination (DR), the subject of this paper, is usually the dominant
recombination process for heavy elements in photoionized nebulae.
The graduate text \citet{2006agna.book.....O} gives an overall introduction to
the physics of photoionized clouds, while
\citet{FerlandSavin01}, 
\citet{Ferland.G03Quantitative-Spectroscopy-of-Photoionized-Clouds},
\citet{Savin.D12The-impact-of-recent-advances-in-laboratory}, and 
\citet{drproject}
review the data needs and the DR process.
\citet{Nikolic.D13Suppression-of-Dielectronic-Recombination-due-to-Finite} 
show how zero-density total DR rate coefficients
can be modified to take account of its suppression at finite densities. 
\citet{Ferland.G03Quantitative-Spectroscopy-of-Photoionized-Clouds}
identifies DR and charge exchange as the two most uncertain processes affecting
the spectroscopy of photoionized clouds. 
Here we consider the rates for S$^{2+} \rightarrow$S$^{+}$ 
radiative and dielectronic recombination.

Low-temperature DR occurs through low-lying autoionizing resonances 
\citep{Nussbaumer.H83Dielectronic-recombination-at-low-temperatures}
 and
is particularly sensitive to their position via the exponent of the Maxwellian distribution.
However, it is inherently difficult to calculate their energy position 
relative to threshold accurately enough and, indeed, whether they reside 
above or below threshold in the first place.
As a result of this ``threshold straddling'' effect, 
\emph{ab initio} theory cannot predict the near threshold DR resonance strength
precisely enough and, ideally, theoretical efforts 
should be guided by experimentally observed positions of autoionizing states in order 
to produce accurate DR rate coefficients at all temperatures \citep{schippers2004}. 
This is especially important for photoionized clouds because the gas kinetic temperature is low 
and so the rate is strongly affected by the position of any such low-lying resonances.
Such experimental energies are not available for S$^+$  so uncertainties in the
positions of the resonances are a source of error,
although estimates have been made
\citep{Badnell.N91Dielectronic-recombination-rate-coefficients, 
Nahar.S95Electron-Ion-Recombination-Rate-Coefficients}.
Although these uncertainties are widely understood, 
we know of no calculations which demonstrate them in detail.
We present such calculations in Section 3 below.

Simulations of nebulae can be used to infer the existence of overlooked processes,
and estimate rates for dominant reactions.
For example,
\citet{Pequignot.D78Charge-transfer-reactions---A-consistent}  inferred 
the existence of fast charge-exchange reactions between atomic hydrogen 
and doubly ionized oxygen from models of a planetary nebula.
In the next section we use the 
\citet{1991ApJ...374..580B} %(hereafter BFM) - defined again at the start of the next section (better?)
model of the Orion Nebula, together with today's advanced stellar atmosphere 
spectral energy distributions (SEDs) and
atomic data, to derive an improved photoionization model.
We reproduce the observed
[O~I], [O~II], [O~III], and [Ne~III] spectra,
produced by ions for which accurate atomic data is available, and
derive an empirical S$^{2+} \rightarrow$S$^{+}$ DR rate coefficient at $T\sim 10^{4}$~K.
Section 3 presents new calculations of the 
 S$^{2+} \rightarrow$ S$^+$ radiative and dielectronic recombination rate coefficients, 
examines the sensitivity of the DR rate coefficient to the detailed
atomic structure, obtains a theoretical rate coefficient
which agrees with the empirical rate coefficient at  $10^{4}$~K, and
investigates its uncertainty over a broader temperature range.
This bootstrap approach can be used to derive DR rate coefficients for other
species using observations of  [O~I], [O~II], [O~III], and [Ne~III].

\section{Bootstrapping S DR from second-row elements in the Orion H II region}

\citet{1991ApJ...374..580B} (hereafter BFM) used Cloudy to construct a 
photoionization model of the Orion Nebula. 
They reproduced the observed intensities of
roughly two dozen emission lines that form in the H$^+$ layer, 
including lines\footnote{
Both [O II] $\lambda$3727 and [S II] $\lambda$6725 are the sum of the 
intensities of the close doublet. } of [O II] $\lambda$3727, [O III] $\lambda\lambda$4363, 5007, 
[S II] $\lambda$6725 and [S III] $\lambda$9532
We improve upon this model here.

\subsection{Improved atomic and grain physics}

We use the most recent stellar SEDs and atomic data, as described below.
The BFM model, and differences between today's assumptions and those
used in 1991, are described here.
We use  version 13.03 of Cloudy,  last described by 
\citet{2013RMxAA..49..137F}.
This has a nearly complete revision of the atomic
database in the 20+ years since BFM was completed.
Some details are given in 
\citet{Ferland.G98CLOUDY-90:-Numerical-Simulation-of-Plasmas} and
\citet{2013RMxAA..49..137F}.
Nearly all aspects of the database have changed,
but improvements in the physics of dielectronic recombination have been substantial,
as outlined below.

BFM included a self-consistent treatment of grains, 
including gas heating by photoejection
of electrons, grain heating by both the stellar continuum
and internally generated radiation, and collisions between grains and the
surrounding plasma.
The grain emission was predicted and found to be in good agreement with observations,
showing that most of the warm dust emission originates in the H$^+$ region.
Our grain treatment has been updated, as described in
\citet{van-Hoof.P04Grain-size-distributions-and-photoelectric} and \citet{2013RMxAA..49..137F}.
As discussed in the first reference, these changes predict less photoelectric heating 
than was found with the older theory used by BFM.
In our current calculation each grain population is resolved into ten sized bins.
Two grain types,
an astronomical silicate and graphitic material, are included.
The size distribution is altered as described by BFM to reproduce the relatively
grey opacities observed in Orion.

\subsection{Geometry}

The BFM model was of a hydrostatic plane-parallel layer on the surface of the
background molecular cloud.  
The ionized gas was assumed to be in hydrostatic equilibrium
(termed ``constant pressure'' in that paper, since all forces balanced)
with the outward momentum of the absorbed stellar continuum being balanced
by the sum of gas and line radiation pressure.
We also assume hydrostatic equilibrium here.
A detailed model of the geometry of the H~II region was derived
by \citet{Wen.Z95A-three-dimensional-model-of-the-Orion-Nebula}.

BFM worked in terms of the flux of ionizing photons striking the plane-parallel layer,
using Cloudy in its so-called ``intensity case'', in which only the flux of photons is specified.
The distance of the stars can be derived from this flux and an assumed calibration
of the O-star luminosity function.
We adopt the stellar parameters and SED described below, which are significantly different
from those assumed in BFM, and reflect advances in our understanding of O stars.
For these parameters, the physical separation between 
the Trapezium stars and the illuminated face
of the H$^+$ layer ($3.15\times 10^{17}$~cm)
 is not  large compared with the thickness of the layer 
 ($6\times 10^{16}$~cm) so the stellar radiation field will suffer some spherical dilution 
as it passes across the layer.
This is taken into account by using Cloudy's ``luminosity case'', in which
the stellar luminosity and star-cloud separation are specified.
This affects the intensities of low-ionization lines.

\subsection{Stellar SED}

The use of modern stellar SEDs is the largest source of  differences between
the calculations presented here and those of BFM.
The SED of the central star cluster is fundamental
because it is the source of heating and ionization in the H$^+$ layer.
BFM used the \citet{Kurucz.R79Model-atmospheres-for-G-F-A-B-and-O-stars}
plane parallel LTE SEDs, 
the most extensive grids of stellar
atmosphere calculations available at that time.
These are now known to produce too soft a radiation field
\citep{Sellmaier.F96A-possible-solution-for-the-Ne-III-problem-in-HII-regions.}.
Several modern stellar atmosphere grids, including TLUSTY 
\citep{Lanz.T03A-Grid-of-Non-LTE-Line-blanketed-Model} 
and WMBasic
\citep{2001A&A...375..161P}, are now available.
These are thought to give a better representation of the SED at ionizing energies
\citep{Simon-Diaz.S08The-ionizing-radiation-from-massive}.
The spectral classes given by
\citet{Simon-Diaz.S06Detailed-spectroscopic-analysis-of-the-Trapezium}
are used.

\begin{figure}[t]
\includegraphics[scale=0.7]{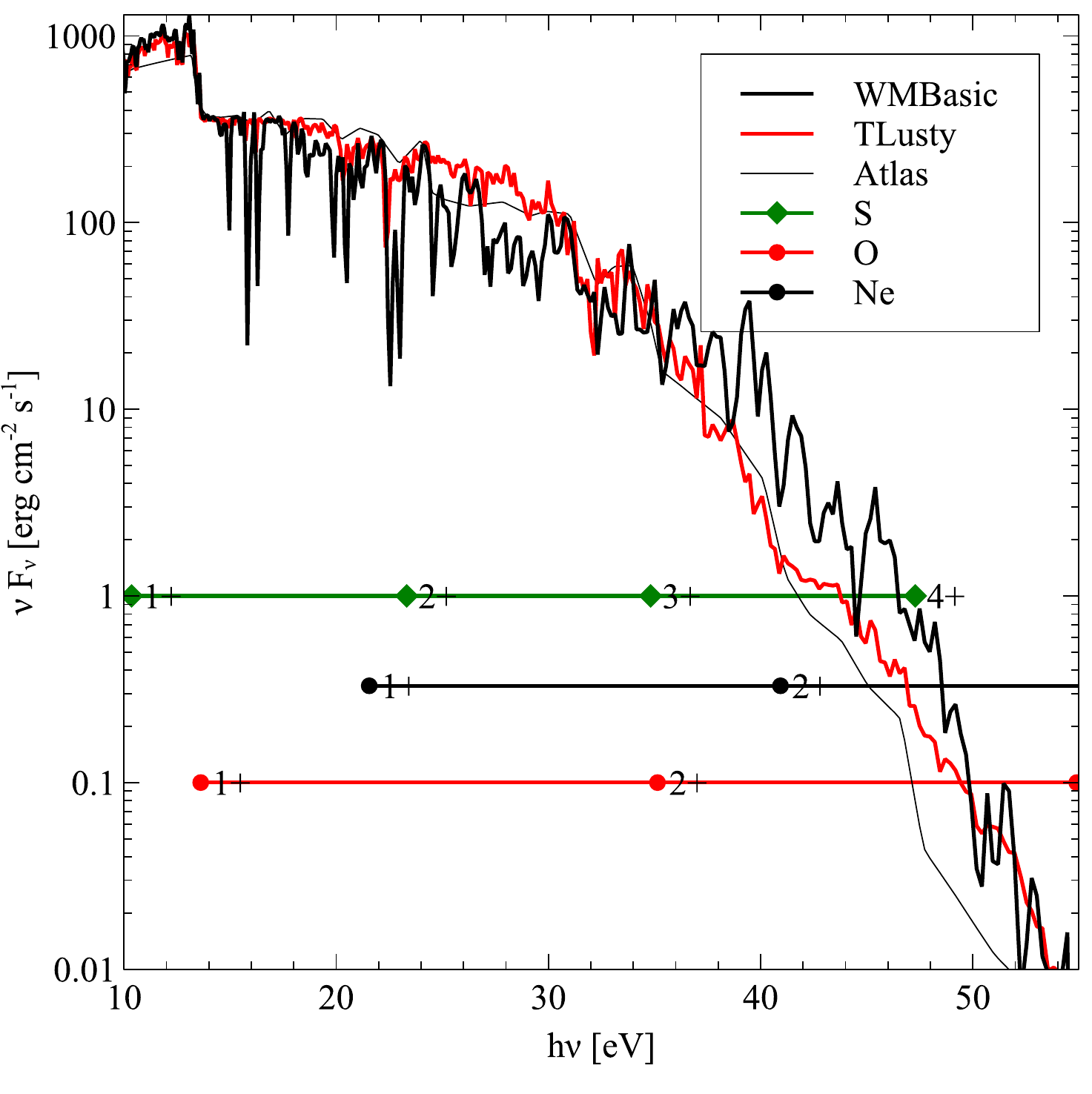}
\caption[Stellar SED and ions of interest]{
Stellar SED and ions of interest.  The Atlas SED was used by BFM while
this work employs WMBasic predictions.
The SED predicted by TLUSTY is also shown.
The horizontal lines indicate ionization potentials for the O, Ne, and S ions
discussed in this paper.
\label{fig:SEDip}}
\end{figure}

Figure \ref{fig:SEDip} compares the Atlas (used by BFM), WMBasic, and TLUSTY SEDs
for the same stellar temperature and luminosity.
The major differences are at photon energies $h\nu \geq 35$~eV,
with the modern calculations  $\sim 1$~dex brighter than Atlas.
This solves a puzzle found by BFM.
They required a high Ne abundance to reproduce the optical
[Ne~III] lines.
[Ne~III] is the highest  ionization species seen in the spectrum of an H II region.
For reference, Figure \ref{fig:SEDip} also shows the ionization potentials of the ions discussed 
in this paper.
O stars have little radiation with h$\nu > 54$~eV.
The high Ne abundance derived by BFM compensated for the deficiency of
photons at energies capable of producing Ne$^{2+}$ in the Atlas SED.
A cosmic Ne abundance is obtained when the modern SED is used, as discussed below.

% data for this Fig in gary dropbox 
% /Users/gary/Dropbox/papers/Badnell_DR/S\ III\ DR/orion/SED_ip

We adopted the WMBasic grid of SEDs.  
These include NLTE and 
the effects of mass loss and winds 
\citep{Pauldrach.A98Realistic-Models-for-Expanding-Atmospheres}.
All are important in OB stars.
\citet{Simon-Diaz.S08The-ionizing-radiation-from-massive}
compare various SEDs and find that WMBasic reproduces the ionization of
H~II regions, and that it is preferred above 35~eV for supergiants with
effective temperatures $\sim40$~kK, which have parameters similar to the stars in Orion.

\subsection{Final model parameters}

Wagle et al (in preparation) describe the updated Orion Nebula model in detail.
Most aspects of the improved model are similar to BFM.
Table \ref{tab:OrionParameters} summarizes model parameters.
The first entry gives the temperature of $\theta^1$ Ori C,
the brightest star in the Trapezium cluster.
The flux of hydrogen-ionizing photons striking the nebula is denoted by $\phi$(H).
The abundances for the interstellar medium (ISM) described in 
\citet{Jenkins.E09A-Unified-Representation-of-Gas-Phase-Element} were used as a reference. 
The abundances are similar, as expected for a newly formed H II region.
Table~\ref{tab:OrionPredictions} gives some predictions, derived using the DR
rate coefficients determined below.  
Most lines are well fitted by the model.
The Table also illustrates some problems with astronomical observations.
The model fits the  [O~III] $\lambda 5007$\AA\  and
[O~II] $\lambda 7325$\AA\ lines quite well, and fits 
 [O~II] $\lambda 3727$\AA\ within 2$\sigma$.
 The UV line is especially uncertain because it lies on the extreme short wavelength edge of the
 spectrum observed by BFM.

%===================================================
\begin{table}
\centering
\caption{\label{tab:OrionParameters}  Model parameters for the Orion Nebula.}
%these change font size making smaller than default
% scope is only in table so does not need to be redone
\null\smallskip
\footnotesize
\renewcommand\arraystretch{0.65}
\begin{tabular}{ c c }
\hline \hline
Parameter & value \\
\hline
$T_*(\theta^1 {\rm Ori C})$ & 38,950 K\\
Radius & 3.15\e{17} cm \\
$\phi$(H)  &  5.9\e{12} cm$^{-2}$ \\
O/H  & 3.66\e{-4} \\
Ne/H  &  1.29\e{-4} \\
S/H   &  1.36\e{-4} \\
\hline
\end{tabular}
\end{table}
%===================================================

%===================================================
\begin{table}
\centering
\caption{\label{tab:OrionPredictions}  Model predictions for the Orion Nebula.}
%these change font size making smaller than default
% scope is only in table so does not need to be redone
\null\smallskip
\renewcommand\arraystretch{0.65}
\begin{tabular}{ c c c }
\hline \hline
Line & Observed & Predicted / Observed \\
\hline
S(H$\beta$) \erg\, cm$^{-2}$ s$^{-1}$ arcsec$^{-2}$	   & 4.63\e{-12} $\pm$0.4 & 1.02 \\
$[$O II$]$	 3727/H$\beta$    & 0.94$\pm$0.2	& 0.64	\\
$[$Ne III$]$     3869/H$\beta$  &	0.2$\pm$0.03	& 0.85	\\
$[$O III$]$      5007/H$\beta$   &3.43$\pm$0.17	& 1.06	\\
$[$S II$]$	6720/H$\beta$     &0.051$\pm$0.003	& 1.01	\\
$[$O II$]$ 	7325/H$\beta$ &0.1191$\pm$0.006	& 0.88   \\
$[$S III$]$	9530/H$\beta$ 	&1.445$\pm$0.285		& 1.00	\\
\hline
\end{tabular}
\end{table}
%===================================================

\subsection{The [Ne~III]/[O~III]/[O~II]/[O~I] bootstrap}

Figure \ref{fig:SEDip} suggests that oxygen / neon spectra can be used to
infer rates for S$^{2+} \rightarrow $S$^+$ recombination.
Second-row elements have relatively complete spectroscopic data,
accurate recombination rate coefficients, and are readily available
(e.g. http://amdpp.phys.strath.ac.uk/tamoc/DR/).
The  observed oxygen ions are produced by the stellar SED between
13.5 eV and 54 eV, while [Ne~III] is produced by photons  between
40 eV and 54 eV.
The relative distribution of these four ions depends on the shape of the 
stellar SED (Figure \ref{fig:SEDip}) and on the intensity of ionizing
radiation striking the layer, as described in the
text \citet{2006agna.book.....O}.

The problem is overdetermined --- we have two variables and three ionization ratios,
counting only second-row ions with good atomic data.
This would guarantee that the [S~III] / [S~II] ratio is properly reproduced since,
as shown by Figure \ref{fig:SEDip}, the S ions lie \emph{within} the
range covered by the O and Ne ions.  In effect, the S ionization
\emph{interpolates} on the observed and reproduced O \& Ne ionization.
The S DR rate is poorly constrained, but the O and Ne ions, with their good
recombination rates, can be used to bootstrap one.

\begin{figure}[t]
\includegraphics[scale=0.5]{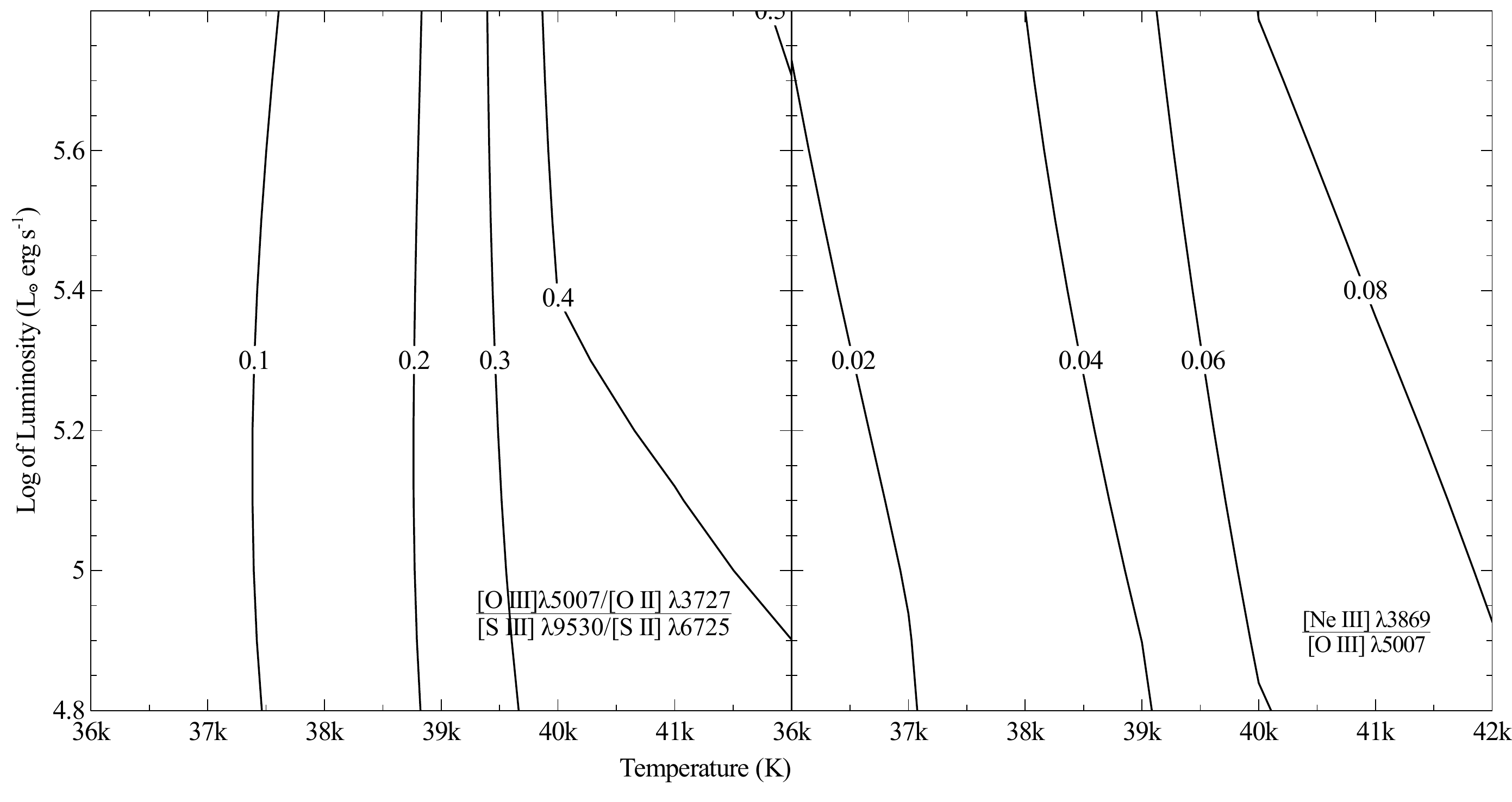}
\caption[The predicted O and Ne spectra]{
Predicted O, S, and Ne spectra are shown as a function of the two free parameters
in the model, the stellar temperature, which sets the SED shape, and the stellar luminosity,
which sets the intensity of starlight falling upon the H$^+$ layer.
The [Ne~III]/[O~III] ratio shown in the right panel sets the stellar temperature, which then
sets the [O~III]/[O~II]~/~[S~III]/[S~II] shown in the left panel.  The S$^{2+}$ recombination
rate is the only additional parameter affecting the left panel.  
\label{fig:Ionization}}
\end{figure}

Figure \ref{fig:Ionization} illustrates this idea.
The right panel shows the ratio of [Ne III] to [O III].
Ne$^{2+}$ has the highest ionization potential of any ion  in an H~II region,
so this ratio is mainly sensitive to the stellar temperature, with little
dependence on stellar luminosity.  The ratio then, in effect,
sets the stellar temperature.

% data for this Fig in gary dropbox 
% /Users/gary/Dropbox/papers/Badnell_DR/S III DR/14_05_30_ms/Ionization.zip Folder

The left panel shows the ratio of ratios, [O~III]/[O~II]~/~[S~III]/[S~II].  
The ratio of ratios depends mainly on
stellar temperature because the O$^{2+}$ -- O$^{+}$ and
S$^{2+}$ -- S$^{+}$ ionization potentials do not match exactly,
so, the fractional abundance ratios change asynchronously.
The stellar temperature is set to good precision by the [Ne~III]/[O~III]. 
The only degree of freedom that can change the left panel is the 
total S$^{2+}\rightarrow $S$^{+}$ recombination rate.

The total S$^{2+}\rightarrow $S$^{+}$ recombination rate is the sum of
a well-determined radiative recombination (RR) rate,
an uncertain dielectronic recombination rate, and possibly charge exchange (CX) recombination.
\citet{Kingdon.J96Rate-Coefficients-for-Charge-Transfer} find that 
S$^{2+}\rightarrow $S$^{+}$ CX does not have an open channel so most
likely occurs by very slow radiative CX, with a rate coefficient $\sim 10^{-17} $ cm$^3$ s$^{-1}$.
We adopt the RR rate coefficient given below, which is in good agreement with that found by 
\citet{Aldrovandi.S73Radiative-and-Dielectronic-Recombination-Coefficients}.
We adjusted the DR rate coefficient to reproduce the observed ratio of ratios
to make Figure \ref{fig:Ionization}.
The resulting best-fit DR rate coefficient is 
 $3 \e{-12}$ cm$^3$ s$^{-1} \pm 0.2$~dex at $10^4$~K, which we refer to as our empirical value.
This kinetic temperature is measured using ratios of [O~III] forbidden lines,
 as outlined in Chapter 5 of \cite{2006agna.book.....O}.
 In this way the well-understood atomic physics of O and Ne can be used
to bootstrap a rate for species lacking the required spectroscopic data.

We  experimented with other SEDs during the course of this work.
We initially used the TLUSTY grid, which is somewhat harder than Atlas but
softer than WMBasic (Figure \ref{fig:SEDip}).
We were able to reproduce the oxygen spectrum using TLUSTY, and simultaneously,
the S$^{2+} - $S$^+$ balance, with a DR rate of  $6 \e{-12}$ cm$^3$ s$^{-1}$.
However, a neon abundance significantly above the cosmic value was needed to
offset the softness of the SED, the same problem that forced a
high neon abundance in BFM.
We prefer to take the neon abundance as a prior, 
which then drives the selection of the SED and the DR rate.

We adopt the solar Ne/H ratio recommended by 
\citet{Asplund.M09The-Chemical-Composition-of-the-Sun}.
This is based on spectroscopic observations of nearby B stars and should
reflect the composition of the local ISM.
Ne is an inert gas and so should not form chemical compounds which then
form grains, so its depletion should be low.
The gas-phase neon abundance of the Orion H II region should be close to the values
obtained from early-type stars.

\begin{figure}[t]
\includegraphics[scale=0.7]{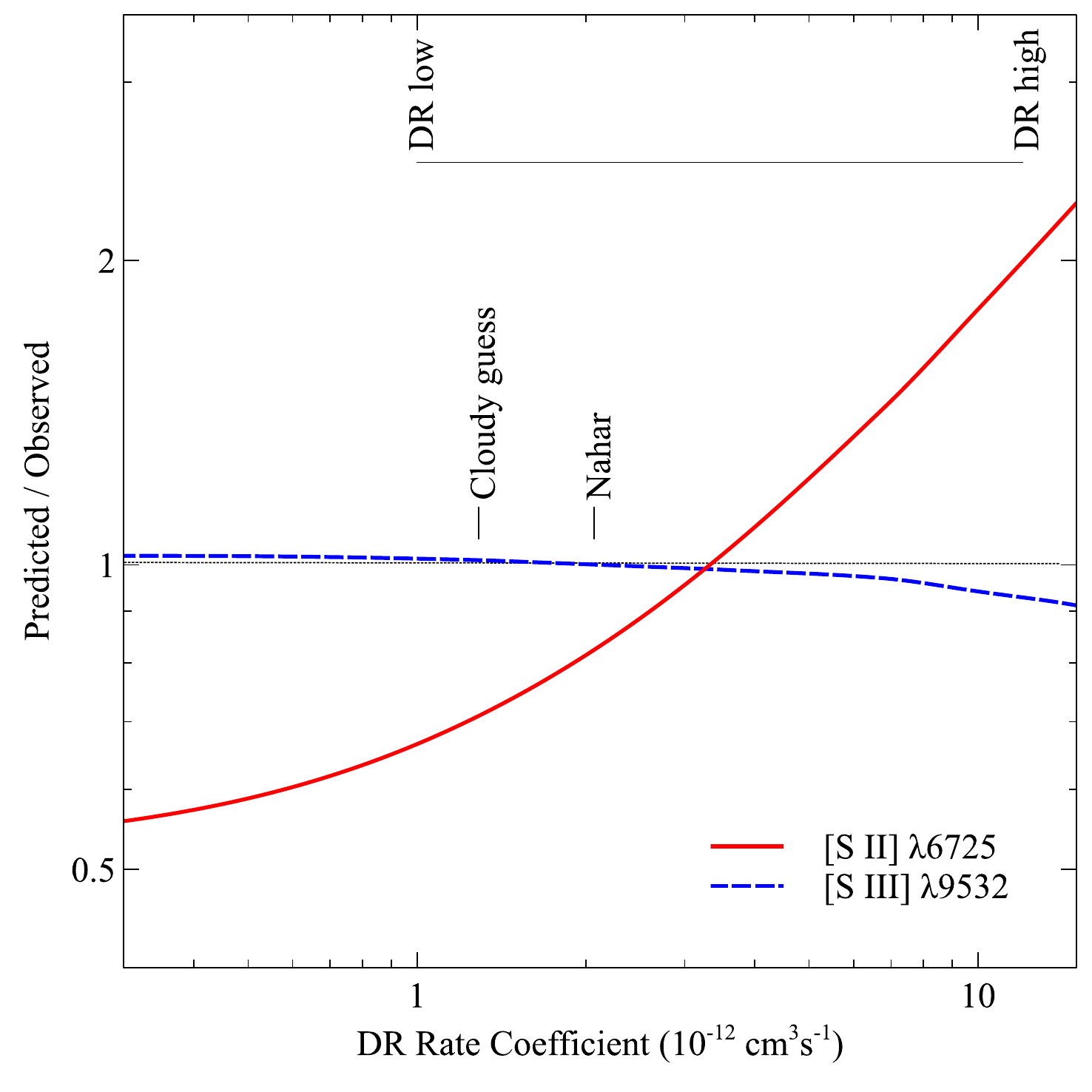}
\caption{
The ratio of predicted to observed line intensities of S lines 
are shown as a function of 
the S$^{2+} \rightarrow $S$^+$ DR rate coefficient.
A DR rate coefficient of $3 \e{-12}$ cm$^3$ s$^{-1} \pm 0.2$~dex is indicated.
\label{fig:OrionSrecomb}}
\end{figure}

Having obtained a best fit DR rate coefficient we can then vary it
to quantify how the line spectrum depends on it.
Strong forbidden lines arising from low-lying levels are predominantly
formed by impact excitation with thermal electrons.  Because of this,
changes in the DR rate change the spectrum mainly through changes in the
fraction of S that is doubly ionized.
This is shown in Figure \ref{fig:OrionSrecomb}.
The main effect of increasing the DR rate coefficient is to increase the intensity of the [S~II] lines.
S$^{2+}$ is the dominant stage of ionization in the H$^+$ region, with only
$\sim 3$\% of S being singly ionized (BFM).  
Increasing the DR rate increases the fraction of S$^+$ significantly while having
a more modest effect on the S$^{2+}$ fraction.
The O lines are hardly affected by changes in the S DR rate.
There are small changes in [O~II] intensities caused by
changes in the gas kinetic temperature as a result of the changing intensities of the
[S II] lines.

Figure  \ref{fig:OrionSrecomb} also shows other estimates of the DR rate coefficient.
\citet{Nahar.S95Electron-Ion-Recombination-Rate-Coefficients} gives the sum
of the RR and DR rate coefficients.
We subtracted the 
\citet{Aldrovandi.S73Radiative-and-Dielectronic-Recombination-Coefficients}
RR rate coefficient to obtain a Nahar ``DR rate coefficient'' of 2.1\e{-12} cm$^3$ s$^{-1}$ at 10$^{4}$~K.
Cloudy uses mean DR rate coefficients for those species not covered by modern calculations,
as described by \citet{Ali.B91The-NE-III-O-II-forbidden-line-spectrum-as-an-ionization}.
The mean for S$^{2+}$ is 1.5\e{-12} cm$^3$ s$^{-1}$ at $10^{4}$~K.
Thus, our empirical total DR+RR rate coefficient is 50\% larger at $10^4$~K.
The atomic calculations described below find values within the range indicated at the top of the figure.

\section{Dielectronic Recombination Calculations for S$^{2+}$ Producing S$^+$}

The previous section has derived a S$^{2+}\rightarrow $S$^{+}$ benchmark DR rate coefficient of
$3 \e{-12}$ cm$^3$ s$^{-1} \pm 0.2$~dex at $10^4$~K.
Here we describe the atomic physics aspects of the DR calculations, and quantify the present 
uncertainties at low temperature associated with low-lying, near-threshold resonances.
The previously-derived rate coefficient is found to be within the rather large
 range of possible computed values. We use that derived benchmark  to constrain our atomic model,
and then compute a consistent DR rate coefficient for all temperatures.

%\subsection{Atomic Description}

The relevant DR and RR processes are as follows. An electron incident on the $3s^23p^2(^3P_0)$ ground 
state of S$^{2+}$ can either directly recombine (i.e., via RR) into any final bound state 
$3s^23p^2(^3P_0)nl$ ($3\le n\le\infty$) or it may first be captured into a particular resonance state 
that subsequently decays radiatively to a final bound state (DR):

{
\begin{eqnarray}
e^- + 3s^23p^2(^3P_{j}) & \longrightarrow & 
\left\{\begin{array}{ll}
3s^23p3dnl & 3p\rightarrow 3d\ {\rm dipole\ resonances}\label{eqa}\\
3s3p^3nl & 3s\rightarrow 3p\ {\rm dipole\ resonances}\label{eqb}\\
3s^23p^2(^3P_{j^\prime})nl & {\rm fine-structure\ resonances}\label{eqac}\\
3s^i3p^j3d^{5-i-j} & (N+1)-{\rm electron\ resonances }\label{eqd}
\end{array}\right. \\
& (RR) \searrow &\ \ \ \ \ \ \downarrow \ (DR) \nonumber \\
& &  \left\{\begin{array}{l}
3s^23p^3 \\
3s^23p^2nl \\
\end{array}\right\}
 + h\nu\ .
	          \label{eqe}
\end{eqnarray}
}

We treat DR and RR independently and proceed to describe the wavefunctions of the initial S$^{2+}$ {\em target}
states, the intermediate  resonance states S$^{+\,**}$, and the final recombined states $3s^23l_13l_2nl$,
where we consider all $n_i=3$ shell orbitals $3l_i=\{3s,3p,3d\}$.
We perform all calculations using the atomic structure and collision code AUTOSTRUCTURE \citep{auto2011},
as applied to numerous DR \citep{drproject} and RR \citep{rrproject} calculations.

\begin{figure}[t]
\includegraphics[width=\linewidth]{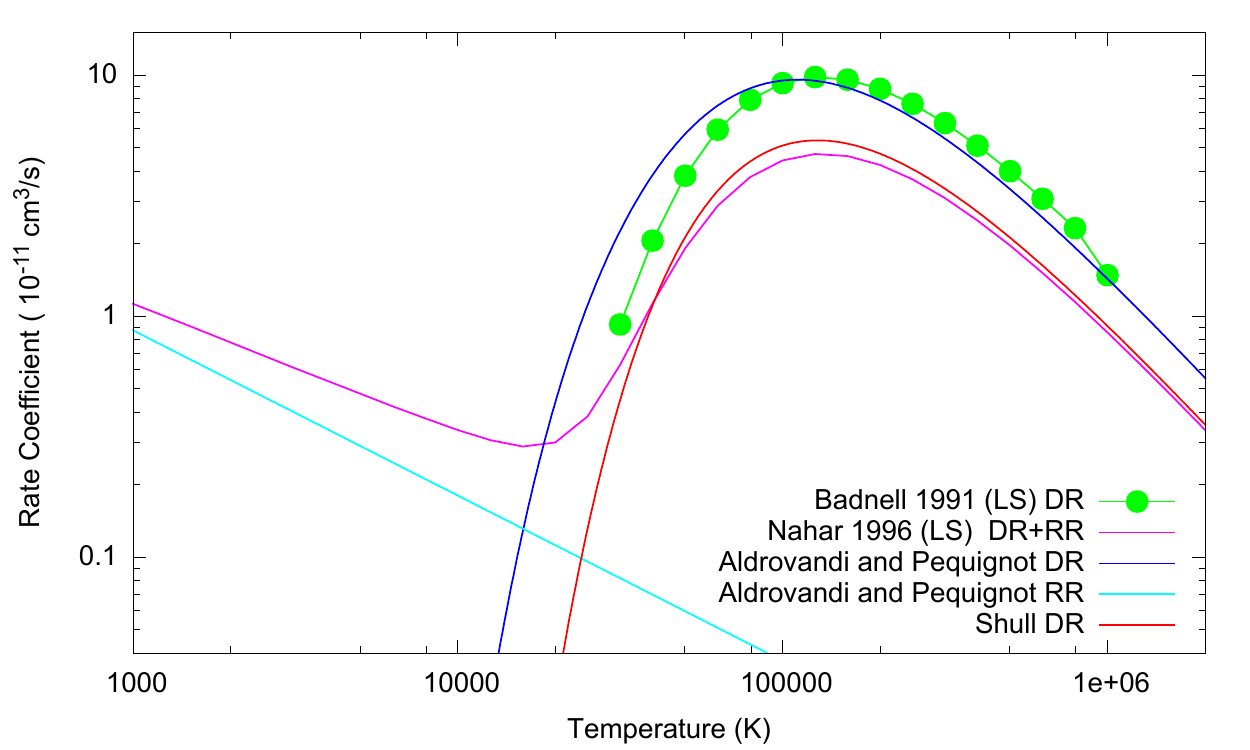}
\caption{Comparison of previously existing S$^{2+}$ DR and RR rate coefficients: 
the LS DR AUTOSTRUCTURE results of \citet{Badnell.N91Dielectronic-recombination-rate-coefficients} (green);
the LS DR-plus-RR R-matrix results of \citet{Nahar.S95Electron-Ion-Recombination-Rate-Coefficients} (magenta);
the DR (blue) and RR (cyan) results of \citet{Aldrovandi.S73Radiative-and-Dielectronic-Recombination-Coefficients};
the DR results of {Shull} (red). 
\label{figprevious}}
\end{figure}

As a means of perspective, we first show the available DR and RR rate coefficient data that was 
available prior to the present study in Figure~\ref{figprevious}.
A prominent high-temperature peak is present
in all the DR results, due to the dipole core-excited contributions in Eq.~\ref{eqa}, but there are minimal features
seen at lower temperatures --- essentially the RR contribution only here. (It is helpful to note that the 
R-matrix results of \citet{Nahar.S95Electron-Ion-Recombination-Rate-Coefficients}
include by necessity the coherent contributions of DR and RR.)
 
Our theoretical treatment for improvement on the existing DR data begins by 
including relativistic effects to first-order via the Breit-Pauli Hamiltonian
(the previous LS calculations of \citet{Badnell.N91Dielectronic-recombination-rate-coefficients} and 
\citet{Nahar.S95Electron-Ion-Recombination-Rate-Coefficients} shown in Fig.~\ref{figprevious} were non-relativistic)
and follows similar recent work on DR of the isoelectronic Fe$^{12+}$ \citep{badnell3pq} and the isonuclear S$^{3+}$ 
\citep{allike} ions.
As in \citet{badnell3pq}, the atomic basis used to describe the S$^{2+}$ target states consists of 
Thomas-Fermi orbitals $\{1s,2s,2p,3s,3p,3d\}$ with 
the same scaling argument $\lambda=1.13$ for all orbitals, which was chosen so as to reproduce the fine-structure 
splitting of the $3s^23p^2(^3P_j)$ levels as compared to NIST (see Table~\ref{tableenergies}).  

%===================================================
\begin{table*}[!h]
\centering
  \caption{  \label{tableenergies} Energies (Ryd) of the S$^{2+}$ target states.} 
  \vspace*{0.07in}
  \begin{tabular}{l l l}  % 3 columns
  \hline \hline
  Level               & Present & NIST\\
  \hline
  % State                   Present    OTHER
 $3s^23p^2\,(^3P_0)$ &   0.000000  &	0.000000   \\[-0.06in]
 $3s^23p^2\,(^3P_1)$ &   0.002712  &	0.002722   \\[-0.06in]
 $3s^23p^2\,(^3P_2)$ &   0.007638  &	0.007591   \\[-0.06in]
 $3s^23p^2\,(^1D_2)$ &   0.128824  &	0.103180   \\[-0.06in]
 $3s^23p^2\,(^1S_0)$ &   0.268614  &	0.247509   \\[-0.06in]
 $3s3p^3  \,(^5S_2)$ &   0.469658  &	0.534658   \\[-0.06in]
 $3s3p^3  \,(^3D_1)$ &   0.747518  &	0.765640   \\[-0.06in]
 $3s3p^3  \,(^3D_2)$ &   0.747707  &	0.765890   \\[-0.06in]
 $3s3p^3  \,(^3D_3)$ &   0.748146  &	0.766370   \\[-0.06in]
 $3s3p^3  \,(^3P_2)$ &   0.888483  &	0.899833   \\[-0.06in]
 $3s3p^3  \,(^3P_1)$ &   0.888986  &	0.900021   \\[-0.06in]
 $3s3p^3  \,(^3P_0)$ &   0.889175  &	0.900078   \\[-0.06in]
 $3s^23p3d\,(^1D_2)$ &   0.944683  &	0.949173   \\[-0.06in]
 $3s^23p3d\,(^3F_2)$ &   1.101637  &	1.112826   \\[-0.06in]
 $3s^23p3d\,(^3F_3)$ &   1.104230  &	1.115427   \\[-0.06in]
 $3s^23p3d\,(^3F_4)$ &   1.107787  &	1.119023   \\[-0.06in]
 $3s3p^3  \,(^1P_1)$ &   1.298494  &	1.247012   \\[-0.06in]
 $3s3p^3  \,(^3S_1)$ &   1.307283  &	1.258155   \\[-0.06in]
 $3s^23p3d\,(^3P_0)$ &   1.323255  &	1.303996   \\[-0.06in]
 $3s^23p3d\,(^3P_1)$ &   1.325432  &	1.304182   \\[-0.06in]
 $3s^23p3d\,(^3P_2)$ &   1.326393  &	1.304254   \\[-0.06in]
 $3s^23p3d\,(^3D_1)$ &   1.357430  &	1.344589   \\[-0.06in]
 $3s^23p3d\,(^3D_2)$ &   1.358329  &	1.345870   \\[-0.06in]
 $3s^23p3d\,(^3D_3)$ &   1.359080  &	1.346358   \\[-0.06in]
 $3s3p^3  \,(^1D_2)$ &   1.440140  &	1.384930   \\[-0.06in]
 $3s^23p3d\,(^1F_3)$ &   1.460029  &	1.436251   \\[-0.06in]
 $3s^23p3d\,(^1P_1)$ &   1.551177  &	1.495763   \\[-0.06in]
  \hline 
  \end{tabular}
\end{table*}
%===================================================

Within this orbital basis, 
we include intra-shell correlation configurations, most importantly, those arising from $3s^2\rightarrow 3p^2$ 
two-electron promotion, giving a target state configuration basis of 
$\{3s^23p^2,3s3p^3,3s^23p3d,3s3p^23d,3s^23d^2,3s3p3d^2,3s3d^3,3p^33d,3p^23d^2,3p3d^3\}$. 
By then performing a large-scale configuration-interaction calculation, including relativistic corrections to lowest order,    
we obtain the S$^{2+}$ target energies listed in Table~\ref{tableenergies}.
While the present computed energies tend to align fairly well with the recommended NIST values, 
especially for the fine-structure-split states, we note theoretical energy overestimates of up to 0.05 Ryd for the higher-lying states.
The dominant high-temperature DR rate coefficient is due to the core dipole transitions 
in Eq.~\ref{eqa}, which are governed by the strongest core radiative rates, and so we list those in Table~\ref{tableradrates}.
The computed rates agree to within $\approx 10\%$ of the NIST values.

%===================================================
\begin{table*}[!h]
\centering  
  \caption{ \label{tableradrates} The three strongest radiative rates $A_r$ ($\times 10^9$ s$^{-1}$) 
from dipole-core-excited states of S$^{2+}$ to the $3s^23p^2\,(^3P_0)$ ground state.} 
  \vspace*{0.07in}
  \begin{tabular}{l c l r r}  % 3 columns
  \hline \hline
  \multicolumn{3}{c}{Transition}          & Present & NIST\\
  \hline
  % State                   Present    OTHER
$3s3p^3  \,(^3S_1)$&$\rightarrow$ & $3s^23p^2\,(^3P_0)$ & 1.82 & 1.60\\[-0.06in]
$3s^23p3d\,(^3P_1)$&$\rightarrow$ & $3s^23p^2\,(^3P_0)$ & 3.51 & 3.87\\[-0.06in]
$3s^23p3d\,(^3D_1)$&$\rightarrow$ & $3s^23p^2\,(^3P_0)$ & 7.27 & 6.93\\[-0.06in]
  \hline 
  \end{tabular}
\end{table*}
%===================================================

\begin{figure}[t]
\vskip -1.0in
\includegraphics[width=0.9\linewidth,height=2.5in]{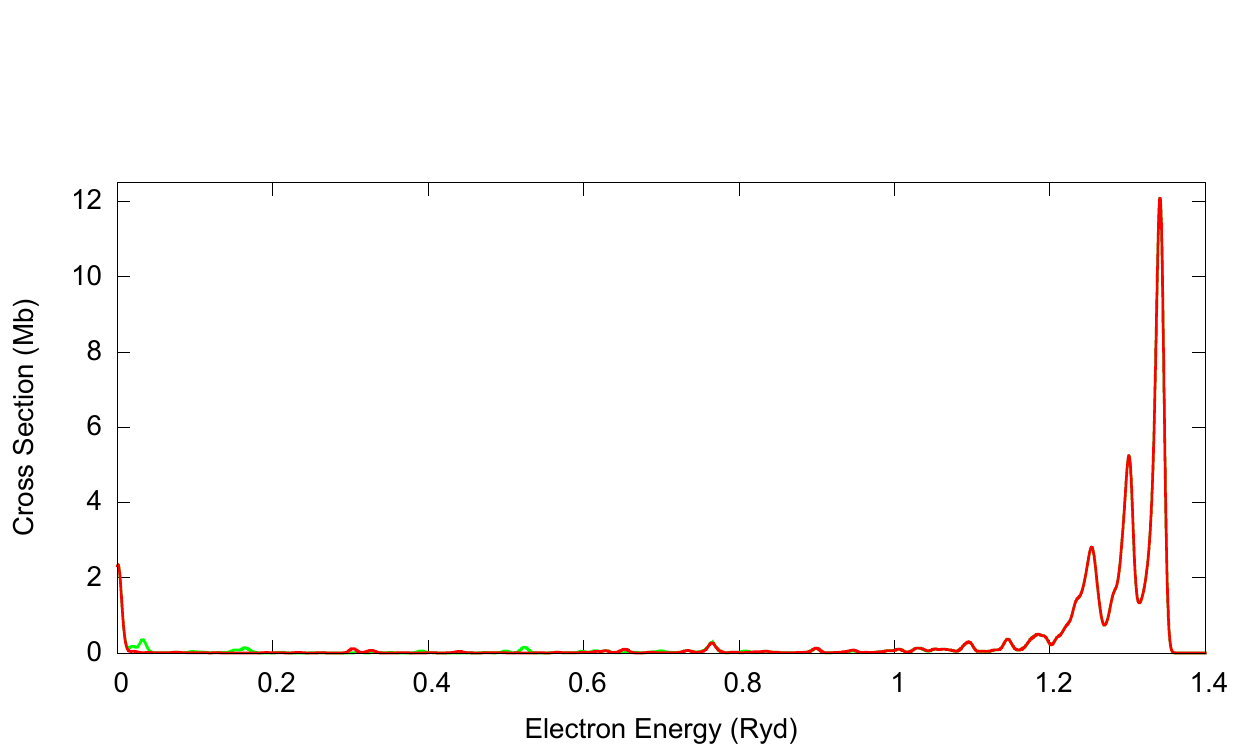} \\[-0.5in]
\includegraphics[width=0.9\linewidth,height=2.5in]{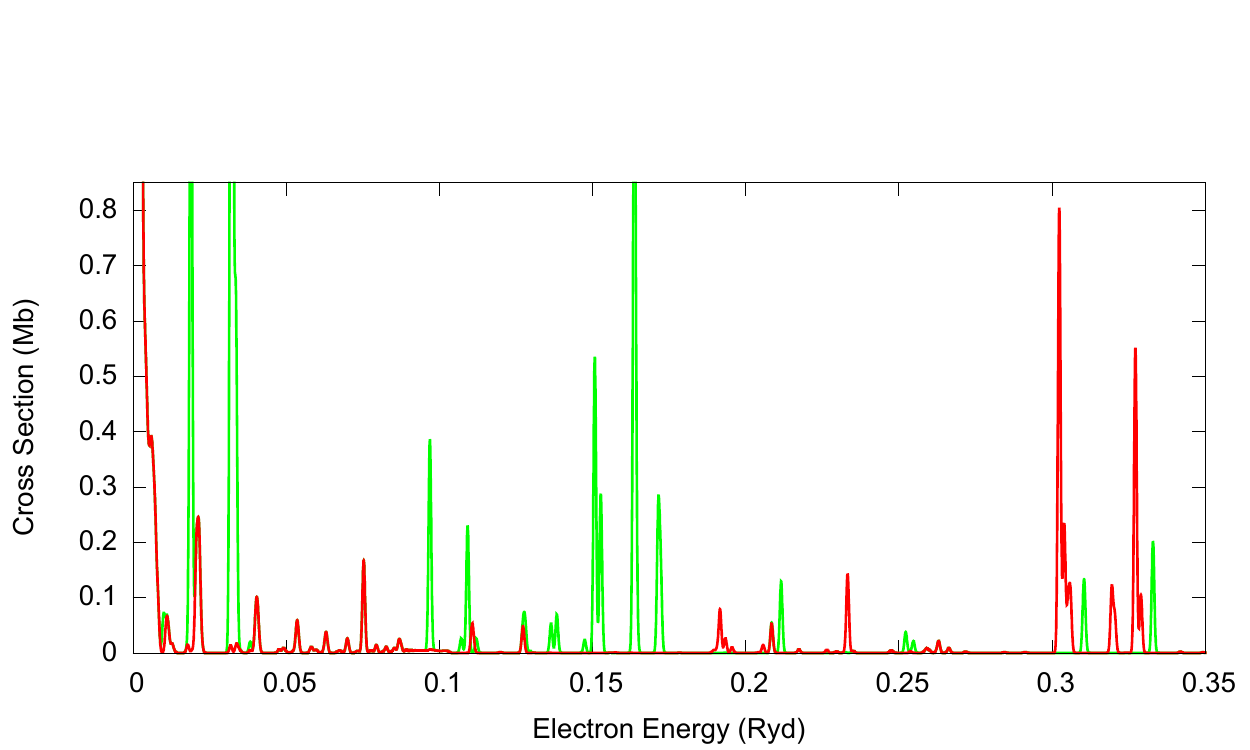} \\[-0.5in]
\includegraphics[width=0.9\linewidth,height=2.5in]{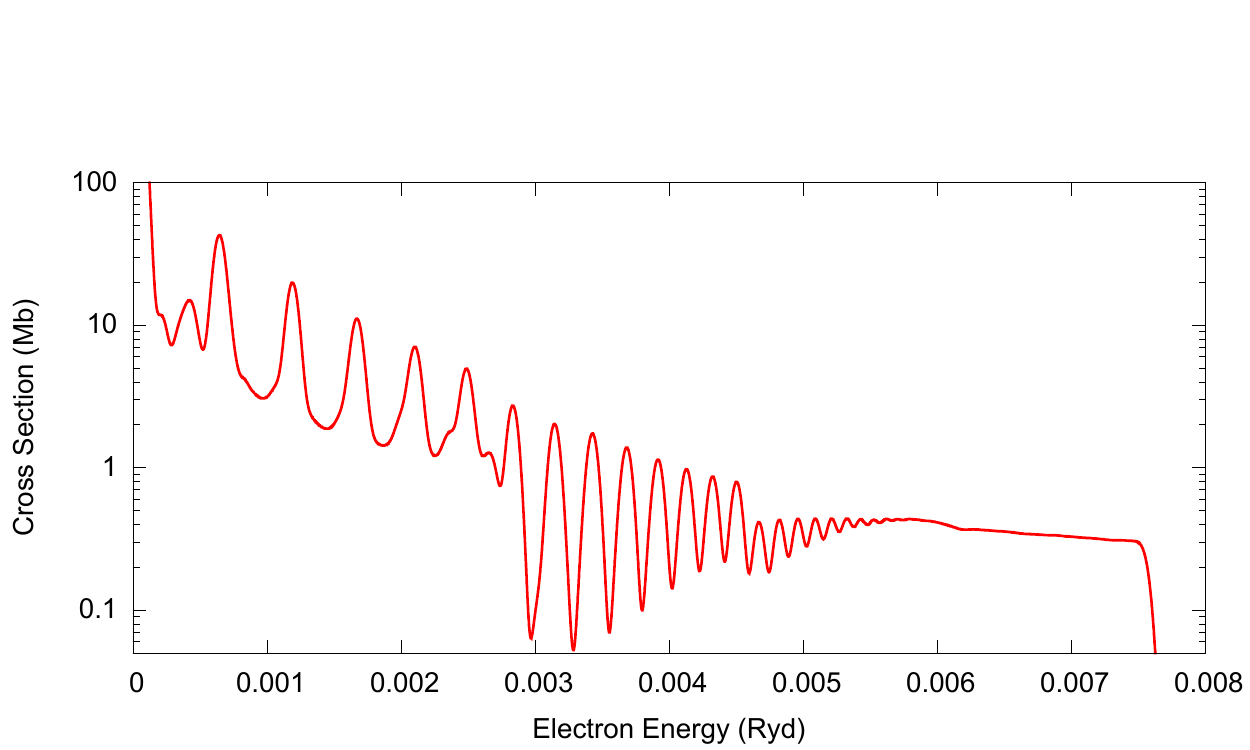} \\[-0.5in]
\includegraphics[width=0.9\linewidth,height=2.5in]{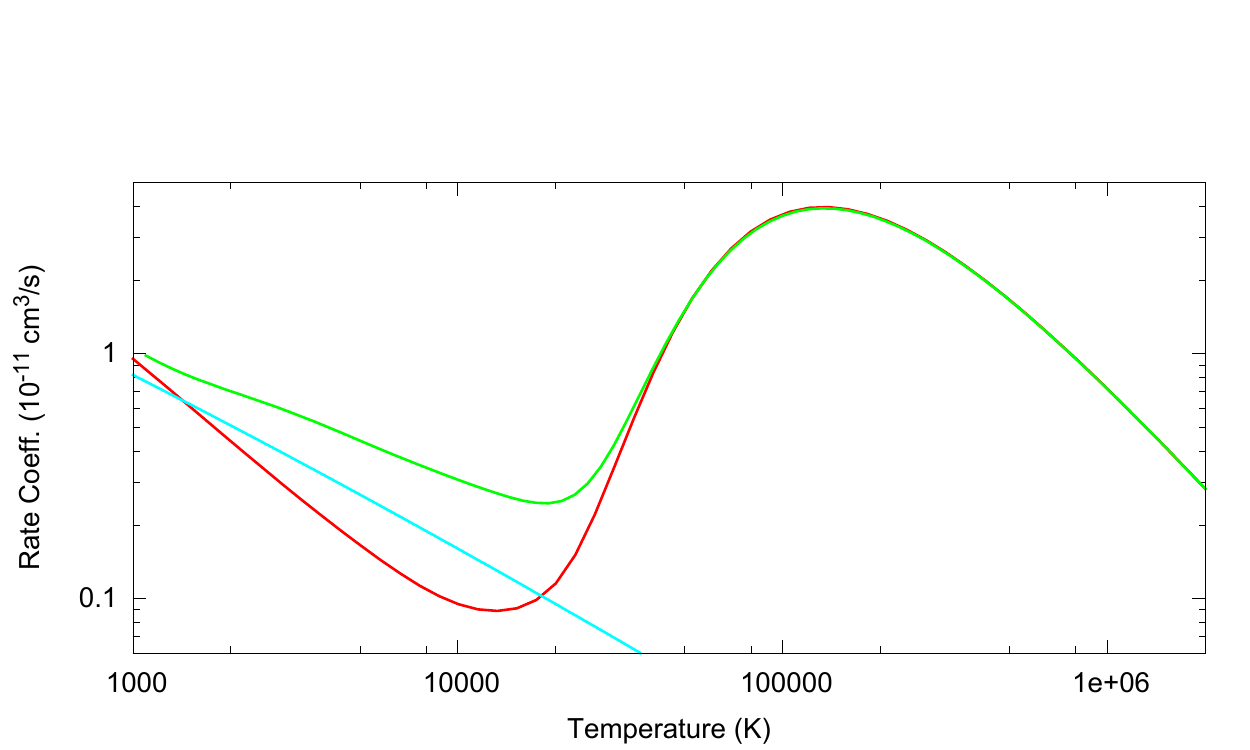} \\[-0.5in]
\caption{S$^{2+}$ DR cross section (upper three panels, convolved with FWHM Gaussians 
convolutions of 0.01, 0.001, and 0.0001 Ryd in descending order) and Maxwellian-averaged
 rate coefficient (lower panel). {\em Ab initio} results are in red and those with all $(N+1)$-electron 
resonances shifted by -0.157 Ryd are in green.  
The computed RR rate coefficient is the cyan curve in the lower panel.
\label{figcross}}
\end{figure}

Having adequately represented the N-electron target states of S$^{2+}$, we then describe the 
various states of S$^{+}$  taking part in Eq.~\ref{eqa}
by a basis consisting of either a distorted-wave continuum ($\epsilon l$) or a valence orbital 
($nl$) coupled to each of the S$^{2+}$ target configurations, plus the so-called $(N+1)$-electron target-orbital basis:
$3s^23p^3$, $3s^23p^23d$, $3s^23p3d^2$, $3s3p^4$, $3s3p^33d$, $3s3p^23d^2$, $3s^23d^3$, 
$3s3p3d^3$, $3p^5$, $3p^43d$, and $3p^33d^2$ (all $N+1=15$ electrons occupy a target orbital 
$nl=\{1s,2s,2p,3s,3p,3d\}$).
Many of these latter $(N+1)$-electron configurations have strong capture and radiative rates, 
and also give rise to near-threshold bound or resonance states,
so they may play an important part in the low-temperature DR process, as will be seen.
Within this large basis set, DR cross sections and Maxwellian-averaged rate coefficients 
are then computed, and these initial results are depicted in Figure~\ref{figcross}.  
The upper panel shows the three strong ground core dipole-excited $3s3p^3(^3S_1)nl$ and 
$3s^23p3d(^3P_1 \,\,\&\,\, ^3D_1)nl$ resonance series 
converging to their respective thresholds ($\approx 1.25-1.35$ Ryd, as indicated in Table~\ref{tableenergies}).  
The second panel from the top focuses
on the low-lying $(N+1)$-electron resonances, in particular the strong $3s3p^33d(^4D_{7/2})$ 
state at 0.321 Ryd (see, also, Table~\ref{tableshift}).
The third panel highlights the $3s^23p^2(^3P_{1,2})nl$ spin-orbit-split series;  these dominate 
the near-threshold energy region (as long as there are no strong low-lying $(N+1)$-electron resonances).  
In the bottom panel, the resultant Maxwellian-averaged rate coefficient indicates that there is a 
large high-temperature peak due to the core dipole-excited series, and then a mild
rise at low temperature due to the spin-orbit-split series near threshold.

It is apparent from Figure~\ref{figcross} that at the temperature of interest --- $10^4$~K,
corresponding to an electron energy of about 0.06 Ryd --- the DR rate coefficient is
determined by the sum of three contributions:
1) the lower-temperature tail of the large contribution from dipole-excited resonances, 2) the higher-temperature 
tail of the near-threshold spin-orbit-split resonances, and
3) the $n=3$ $(N+1)$-electron resonances, with uncertain resonance positions.
We note that while contributions 1) and 2) are subject to suppression at finite densities,
according to \citet{Nikolic.D13Suppression-of-Dielectronic-Recombination-due-to-Finite}, contribution 3) is not.

Upon closer scrutiny of the energies of the $(N+1)$-electron bound states and resonances, 
and by comparing to available NIST bound state data, it is found that our limited (for computational feasibility) 
atomic description results in a general overestimate of these latter energies.
In Table~\ref{tableshift}, we list three prominent $(N+1)$-electron S$^+$ bound state energies ---
for the  $3s3p^4\,(^2S_{1/2})$, $3s3p^4\,(^2P_{3/2})$, and $3s3p^4\,(^2P_{1/2})$ states.
Relative to the S$^{2+}3s^23p^2\,(^3P_0)$ threshold, corresponding to zero incident electron energy 
in the DR process, the theoretically-predicted energies are overestimated by  $\approx 0.25-0.45$ Ryd. 
Also listed in this table is the  strong S$^+\,3s3p^33d\,(^4D_{7/2})$ resonance, which has a theoretically-predicted 
energy position of $\approx 0.32$ Ryd (there are no available NIST or other data for this autoionizing energy position, 
to our knowledge). Given the overestimate of the corresponding  bound $(N+1)$-electron energies, it is reasonable to infer that the 
energy of this autoionizing  resonance is likewise overestimated, and that the ``true'' resonance position should be closer 
to threshold;  this suggests that the ``true''  DR rate coefficient at $T=10^4$~K could be enhanced by the presence
of such a strong resonance, compared to a computed rate coefficient where the theoretical resonance position is at higher energy.
More generally, we expect the entire manifold of $(N+1)$-electron resonances to be positioned too high and
so there may be similar contributions (DR enhancements) from a group of suitably re-positioned resonances.

To understand how the DR rate coefficient at $T=10^4$ depends on the precise position of a strong resonance, 
consider that, for an isolated resonance, the rate coefficient is given by 
\begin{eqnarray*}
\alpha^{{\rm DR}}_{i}(T) & \approx & constant \times  T^{-3/2}{\rm e}^{-E_i/kT}\ ,
\end{eqnarray*}
where $E_i$ is the resonance position.
For a given resonance strength, this contribution at $T=10^4\ (kT\approx 0.06\,{\rm Ryd})$ 
increases as  the resonance position approaches threshold.  For example, the enhancement to the $T=10^4$ rate
coefficient by repositioning, or shifting, a single resonance from $E_i=0.32$ Ryd to $E_i^{shift}\approx 0$  
is given by the factor ${\rm e}^{+E_i/kT}\approx {\rm e}^{0.32/0.06}\approx 200$.
 This example illustrates how uncertainties in resonance positions can translate into rather large
 uncertainties in the {\em low-temperature} DR rate coefficient. (At higher temperatures, any energy uncertainty 
is small compared to $kT$ so that there is far less sensitivity to the exact resonance position.)

%===================================================
\begin{table*}[!h]
  
  \caption{Energies (Ryd) of the S$^{+}$ ground state and selected $(N+1)$-electron bound and resonance states, 
compared to the S$^{2+}$ ground state. The present theoretical energies for $3s$-vacancy states, 
which are subject to relaxation effects, relative to the ground state of S$^{2+}$, are overestimated by $\approx 0.2-0.4$ Ryd., 
as highlighted in boldfaced font. \label{tableshift}} 
  \vspace*{0.07in}
  \begin{tabular}{l@{\hspace{0.1in}}l | r r | r r r}  % 5 columns
  \hline \hline
   &   & \multicolumn{2}{|c|}{$E-E_{3s^23p^3\,(^4S_{3/2})}$} & \multicolumn{2}{c}{$E-E_{3s^23p^2\,(^3P_{0})}$} & \\
 \multicolumn{2}{c|}{Ionic State}             & Present & NIST & Present & NIST& Overestimate\\
  \hline
  % State                   Present    NIST
S$^+$ & $3s^23p^3\,(^4S_{3/2})$ &     0.000 &	0.000    &    -1.630 &	-1.715  &   \\[-0.00in]
S$^+$ & $3s3p^4\,(^2S_{1/2})$   &     1.455 &	1.092    &    -0.175 &	-0.623  &  {\bf +0.448} \\[-0.00in]
S$^+$ & $3s3p^4\,(^2P_{3/2})$   &     1.488 &	1.326    &    -0.142 &	-0.389  &  {\bf +0.247} \\[-0.00in]
S$^+$ & $3s3p^4\,(^2P_{1/2})$   &     1.493 &	1.329    &    -0.137 &	-0.386  &  {\bf +0.249} \\[0.1in]
 \hline
S$^{2+}$ & $3s^23p^2\,(^3P_0)$   &     1.630 &	1.715    &     0.000 &	 0.000  &   \\[0.1in]
 \hline
S$^+$ & $3s3p^33d\,(^4D_{7/2})$ &     1.951 &	         &     0.321 &	        & \\[-0.00in]
  \hline 
  \end{tabular}
\end{table*}
%===================================================

Guided by our initial supposition that the computed low-lying resonance energy positions are overestimated, 
we wish to investigate the effect of resonance position uncertainties on the low-temperature DR rate coefficient
by performing two new DR calculations in which the $(N+1)$-electron resonance positions are shifted.
In the first case, a global $(N+1)$-electron resonance shift of -0.32 Ryd
was applied in order to realign the strongest S$^+\,3s3p^33d\,(^4D_{7/2})$
resonance to just above threshold (at +0.001 Ryd), thereby maximizing the resulting low-temperature rate coefficient.  
In the second case,  a global shift of -0.157 Ryd was used, 
positioning instead that resonance at only 0.163 Ryd above threshold, but
yielding a total DR rate coefficient of $3\times 10^{-12}$ cm$^3$/s at $T=10^4$~K, consistent with the value derived in the previous section.

\begin{figure}[t]
\includegraphics[width=\linewidth]{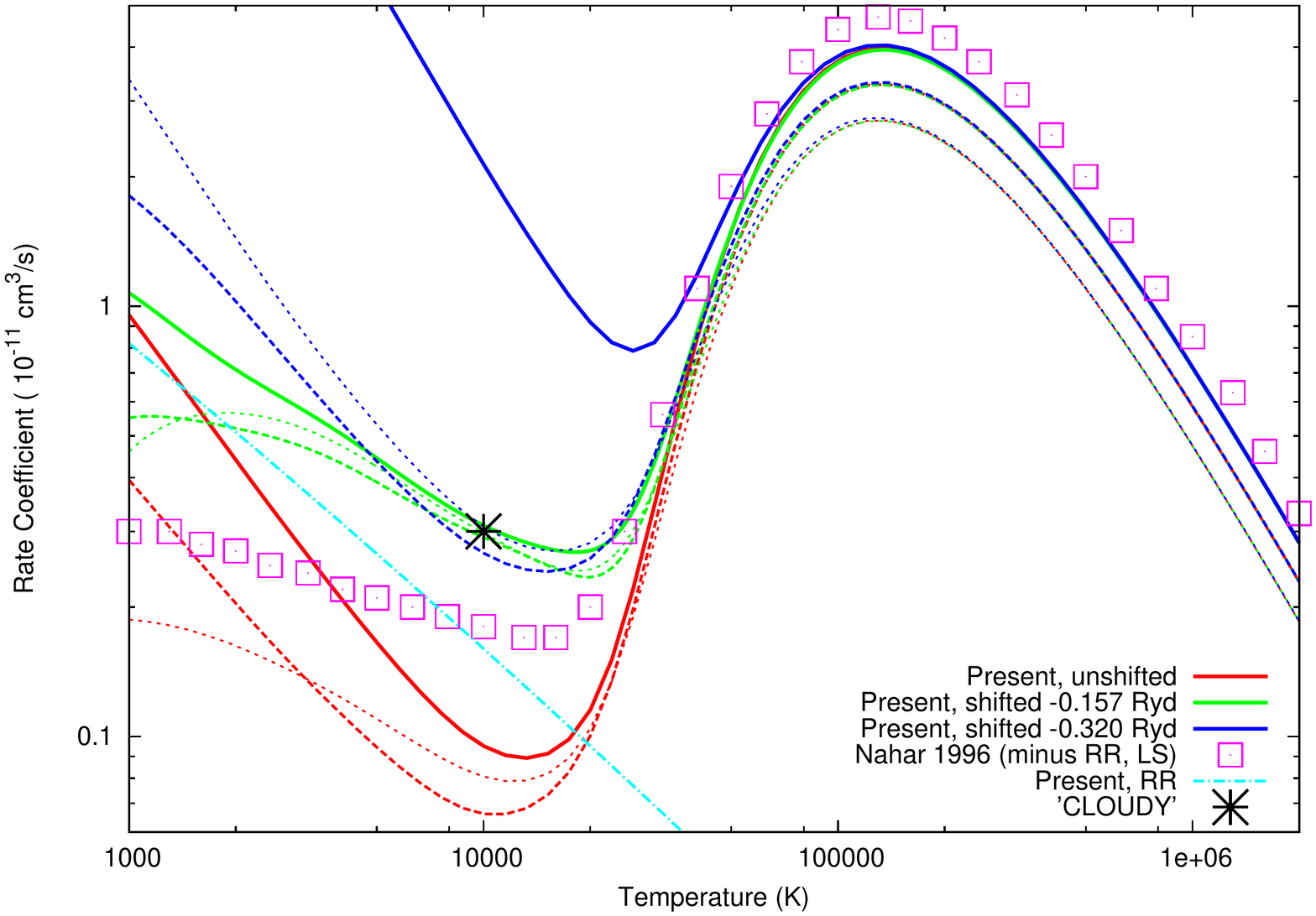}
\caption{
A comparison of S$^{2+}$ DR and RR rate coefficients, including the breakdown of DR from the $3s^23p^2(^3P_j)$ 
ground ($j=0$, solid) and metastable ($j=1$, long-dashed) and ($j=2$, short-dashed) levels.  The red curve indicates 
the \emph{ab initio} calculations, where the strong $3s3p^33d(^4D_{7/2})$ resonance is positioned at 0.321 Ryd above threshold.
The blue curve corresponds to an empirical shift of all $n=3$ $(N+1)$-electron resonances by -0.32 Ryd so as to 
reposition the $3s3p^33d(^4D_{7/2})$ resonances just above the $j=0$ ground state threshold, giving a maximum 
$j=0$ low-temperature rate coefficient.  The green curve represents an artificial shift somewhere in between (-0.157 Ryd) which
yields a statistically-averaged rate coefficient at $T=10^4$~K (see Table~\ref{tabledrrates}) consistent with the Cloudy deduced value
(black asterisk) of $3\times 10^{-12}$ cm$^3$/s.  Also shown are the present RR results (cyan curve) and 
an estimate of the DR contribution from \citet{Nahar.S95Electron-Ion-Recombination-Rate-Coefficients}
(magenta curve).
\label{figcompare}}
\end{figure}

The computed DR rate coefficients are shown in Figure~\ref{figcompare} for all three cases ---  \emph{ab initio} (no shift), 
a shift of -0.32 Ryd, and a shift of -0.157 Ryd --- and each case is further resolved by the various initial 
$3s^23p^2(^3P_j)$ levels, with $j=0$ corresponding to the ground state of S$^{2+}$ and $j=1,2$ being the 
first two spin-orbit-split excited states (see Table~\ref{tableenergies}).  
At lower temperatures, in particular, at $T=10^4$~K, the rate coefficient varies by about an order of 
magnitude between the \emph{ab initio} results and the $-0.32$ Ryd shifted results.
It should be noted, however, that the latter case corresponds to the optimal 
shift scenario in which the strongest resonance was positioned closest to threshold.
Furthermore, this fortuitous positioning of the strong resonance just above the $j=0$ ground state 
(therefore {\em below} the two $j=1$ and $j=2$ excited states) results in a larger enhancement in 
the ground-state rate coefficient compared to 
the excited-state rate coefficients.  In such cases where the various $j$-resolved results differ significantly,  
an observation-based rate coefficient 
could be used in conjunction with the theoretically-predicted $j$ variations to obtain density diagnostics.

%===================================================
\begin{table}
\centering
\caption{Computed S$^{2+}$ DR and RR recombination rate coefficients (cm$^3$ s$^{-1}$)
evaluated at $T = 10^4$ K.  
The computed results are resolved by ground ($j=0$) and metastable  ($j=1$ and $j=2$)  
$3s^23p^2(^3P_{j})$ initial states. The computed statistical averages and the 
high-density limit (HDL) derived rate coefficients are also listed.\label{tabledrrates}}
\null\smallskip
\renewcommand\arraystretch{0.65}
\begin{tabular}{ r c c }
\hline \hline
$j$ & DR & RR \\
\hline
0      & 3.08E-12 & 1.60E-12 \\
1      & 2.82E-12 & 1.59E-12 \\
2      & 2.98E-12 & 1.50E-12 \\
\\
Avg.   & 2.94E-12 & 1.54E-12 \\
\hline
HDL & 3.10E-12 & 1.54E-12 \\
\hline
\end{tabular}
\end{table}
%===================================================

The last case involved an intermediate scenario in which the $(N+1$)-electron resonances were shifted by $-0.157$~Ryd.
This particular shift was chosen so as to produce a statistically-averaged (over initial $j$ levels)
DR rate coefficient in line with the high-density limit (HDL) value of $3\times 10^{-12}$ cm$^3$/s
derived earlier (see Table~\ref{tabledrrates}, which lists the various DR and RR rate coefficients at $T=10^4$~K).
Thus we have {\em bootstrapped} our DR calculations, that were initially suspect
due to uncertainties in the $(N+1$)-electron resonance
positions, by choosing a plausible shift of these resonances in order to reproduce a computed rate coefficient
that agrees with the benchmark value at $T=10^4$~K.  

\begin{figure}[t]
\includegraphics[width=\linewidth]{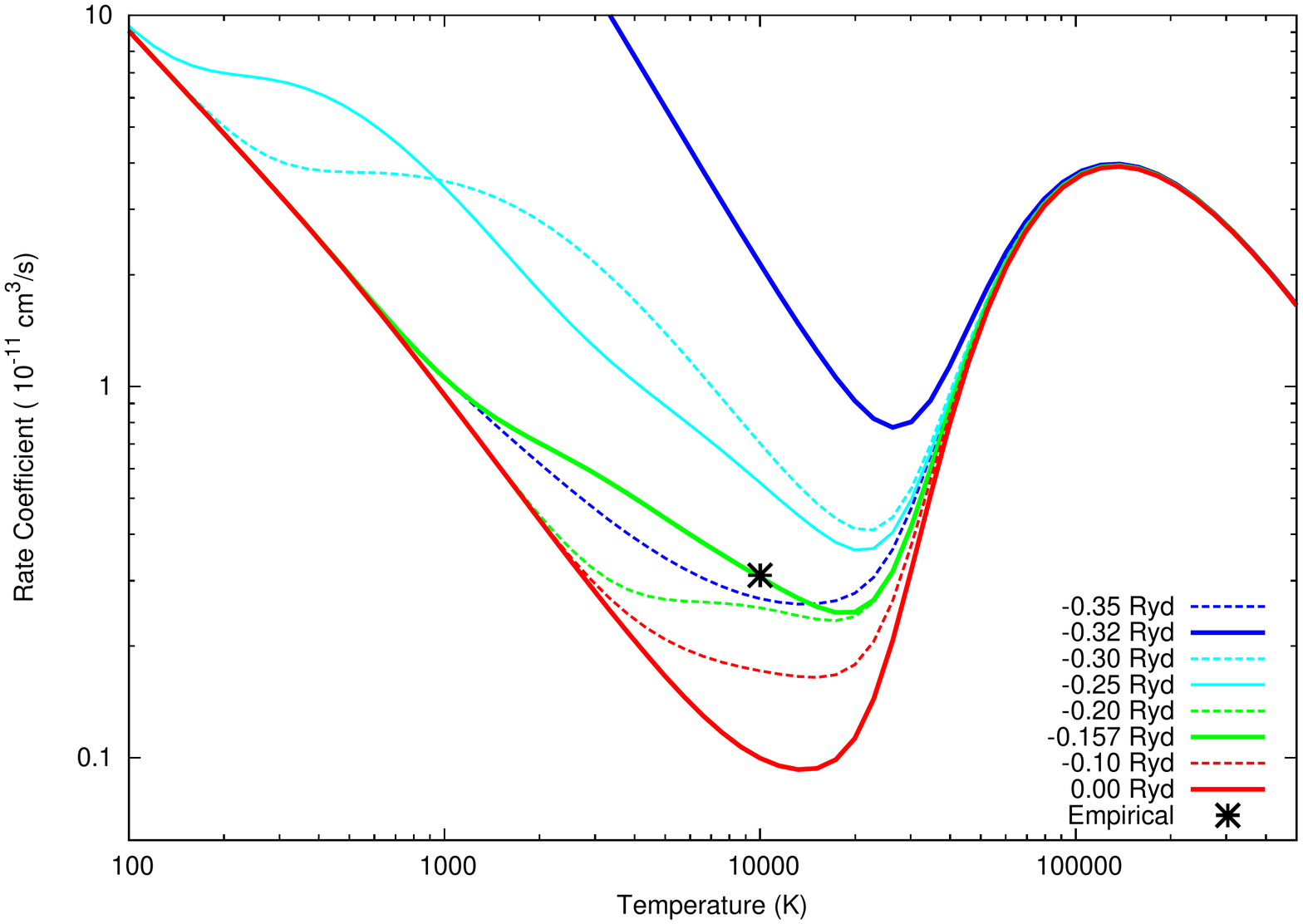}
\caption{DR rate coefficients for electrons incident on the
S$^{2+}(3s^23p^2\,^3P_0)$ ground state, subject to a uniform shift in energy position of all $(N+1)$-electron resonances.  
The shifted energies for each case are shown in the legend, and the present observationally-derived empirical
rate coefficient of $3.10\times 10^{-12}$ cm$^3$/s at $T=10^4$~K is also shown.
\label{figtrend}}
\end{figure}

To extend the usefulness of this approach to other photoionized plasmas (e.g. over $10^3 - 10^4$~K), we need to examine
the spread of any family of curves, corresponding to different shifts, say, but which pass through our point at $10^4$~K.
We do this in Figure~\ref{figtrend}, for the ground level only now. Of particular note are the three curves, 
with shifts of -0.20, and -0.35 Ryd, plus our original -0.157 Ryd. 
While all three are consistent with our empirical
value at $10^4$~K, they show a rather broad variation away from this temperature, particularly at lower $T$.
This is due to the fact that the increase at $10^4$~K, over the unshifted result, is due to the contribution of
several resonances. To reduce the present uncertainty for this ion away from $10^4$~K we either need an
observation at a second (lower) temperature to enable us to fix the slope of theoretical rate coefficient,
as well as its magnitude, or we need to carry-out a more extensive  $(N+1$)-electron structure calculation so
as to try and reduce the uncertainty in position of this manifold of resonances, i.e., reduce the range of plausible
energy shifts.

For all curves consistent with our empirical value, it is the $n=3$ $(N+1)$-electron resonances which dominate
DR at the main temperatures of interest for photoionized plasma. As we have already noted, they should not be 
suppressed by density effects. The work of
\citet{Nikolic.D13Suppression-of-Dielectronic-Recombination-due-to-Finite} classifies the Si-like sequence
as one for which DR suppression applies at/for all temperatures/resonances since it assumes that the
fine-structure peak dominates at low temperatures. For the specific case of S$^{2+}$, we re-classify it
as an ion for which DR suppression is turned-off as the temperature drops below the high-temperature
DR peak. Thus, we now use equation (14) of 
\citet{Nikolic.D13Suppression-of-Dielectronic-Recombination-due-to-Finite} for S$^{2+}$
with $\epsilon = 1.3$~Ryd (=17.7~eV in the units of (14)), this being representative of the dipole core-excitation
energies which give rise to the high-temperature DR peak.

Absent any  additional resonance information, experimental or from more converged atomic structure calculations, 
we choose the final model for computing the DR rate coefficients at all temperatures to be that resulting
from a shift of $-0.157$~Ryd. 
For efficient use in modeling codes such as Cloudy, it can be fit by the expression  
 \begin{eqnarray*}
\alpha^{{\rm DR}}(T) & = &  T^{-3/2}\sum_i c_i{\rm e}^{-E_i/T}\ .
\end{eqnarray*}
The DR fitting coefficients $c_i$ and $E_i$  for each $j$ are
listed in Table~\ref{tabledrfit} and are accurate to better than 5\% above $T=200$~K.
The total RR rate coefficient, which is fairly insensitive to the initial 
state $j$, can be fit by  
 \begin{eqnarray*}
 \alpha^{{\rm RR}}(T) = A \, \sqrt{T_0/T} \, \left[ \left( 1 + \sqrt{T/T_{0}} \right)^{1-B'}
 \left(1 + \sqrt{T/T_{1}} \right)^{1+B'} \right]^{-1} \ ,
\end{eqnarray*}
where
\begin{equation}\label{Eq:Fit:RR-low}
 B' = B + C \exp(-T_2/T)\ ,
\end{equation}
and these coefficients are
listed in Table~\ref{tablerrfit}.

\begin{table*}[!htbp]
%\begin{minipage}[t]{0.48\textwidth}
\caption{\label{tabledrfit} DR fitting coefficients
$c_i$ (in cm$^3$\,K$^{3/2}$\,s$^{-1}$) and $E_i$ (in K) for the $j=0,1,2$ levels of the S$^{2+}$
$^3$P ground term. 
%$\alpha^{{\rm DR}}(T) = \frac{1}{T^{3/2}} \sum_{i} c_{i} \exp\left(- \frac{E_{i}}{T}\right)$
}

\scriptsize
%\resizebox{\textwidth}{!}{
\begin{tabular}{l c c c c c c c}
\hline
\hline
initial state & $c_1$ & $c_2$ & $c_3$ & $c_4$ & $c_5$ & $c_6$ & $c_7$ \\ 
\hline
$e^-+3s^23p^2(^3P_0)$ &
  2.539E-07	 &
  4.184E-07	 &
  2.763E-06	 &
  1.035E-05	 &
  7.592E-05	 &
  8.686E-03	 &
  5.991E-05
                \\[.1in]
$e^-+3s^23p^2(^3P_1)$ &
  1.130E-07    &
  4.769E-07    &
  2.841E-06    &
  1.847E-05    &
  4.040E-04    &
  3.371E-03    &
  3.371E-03   
      \\[.1in]

$e^-+3s^23p^2(^3P_2)$ &
 3.176E-08	 &
 9.676E-07	 &
 2.311E-06	 &
 1.139E-05	 &
 1.645E-04	 &
 5.610E-03	 
 	\\
\hline
\hline
initial state & $E_1$ & $E_2$ & $E_3$ & $E_4$ & $E_5$ & $E_6$ & $E_7$ \\ 
\hline
$e^-+3s^23p^2(^3P_0)$ &
  1.244E+02    &
  1.428E+03    &
  5.411E+03    &
  2.514E+04    &
  8.384E+04    &
  2.050E+05    &
  7.858E+05    
                \\[.1in]
$e^-+3s^23p^2(^3P_1)$ &
 2.185E+02  &
 1.889E+03  &
 5.729E+03  &
 3.255E+04  &
 1.381E+05  &
 2.047E+05  &
 2.090E+05  \\[.1in]
$e^-+3s^23p^2(^3P_2)$ &
   2.384E+02   &
   2.149E+03   &
   5.553E+03   &
   2.616E+04   &
   1.030E+05   &
   2.031E+05	 \\
\hline
\end{tabular}
%}
%\end{minipage}
\end{table*}

\begin{table*}[!htbp]
%\begin{minipage}[t]{0.48\textwidth}
\caption{ \label{tablerrfit} RR fitting coefficients for the ground state of S$^{2+}$. 
%$\alpha^{{\rm RR}}(T) = A \, \sqrt{T_0/T} \, \left[ \left( 1 + \sqrt{T/T_{0}} 
%\right)^{1-B'(T)}\left(1 + \sqrt{T/T_{1}} \right)^{1+B'(T)} \right]^{-1}$,
%$B'=B + Ce^{-T_2/T}$
}
% \resizebox{\textwidth}{!}{
\begin{tabular}{lcccccc}
\hline
\hline
     $A$ (cm$^3$\,s$^{-1}$)   &	$B$ &  $T_{0}(K)$ &  $T_{1}(K)$  & $C$ & $T_{2}(K)$\\
% (cm$^3$s$^{-1}$)&    & (K)	 &  (K) &    & (K)	  \\
\hline
  2.326E-11 & 0.4601 & 3.639E+02 & 2.170E+07 &  0.3398 &  7.597E+05  \\
\hline
\hline
\end{tabular}
%}
%\end{minipage}
\end{table*}

\section{Discussion}

\begin{figure}[t]
\includegraphics[width=\linewidth]{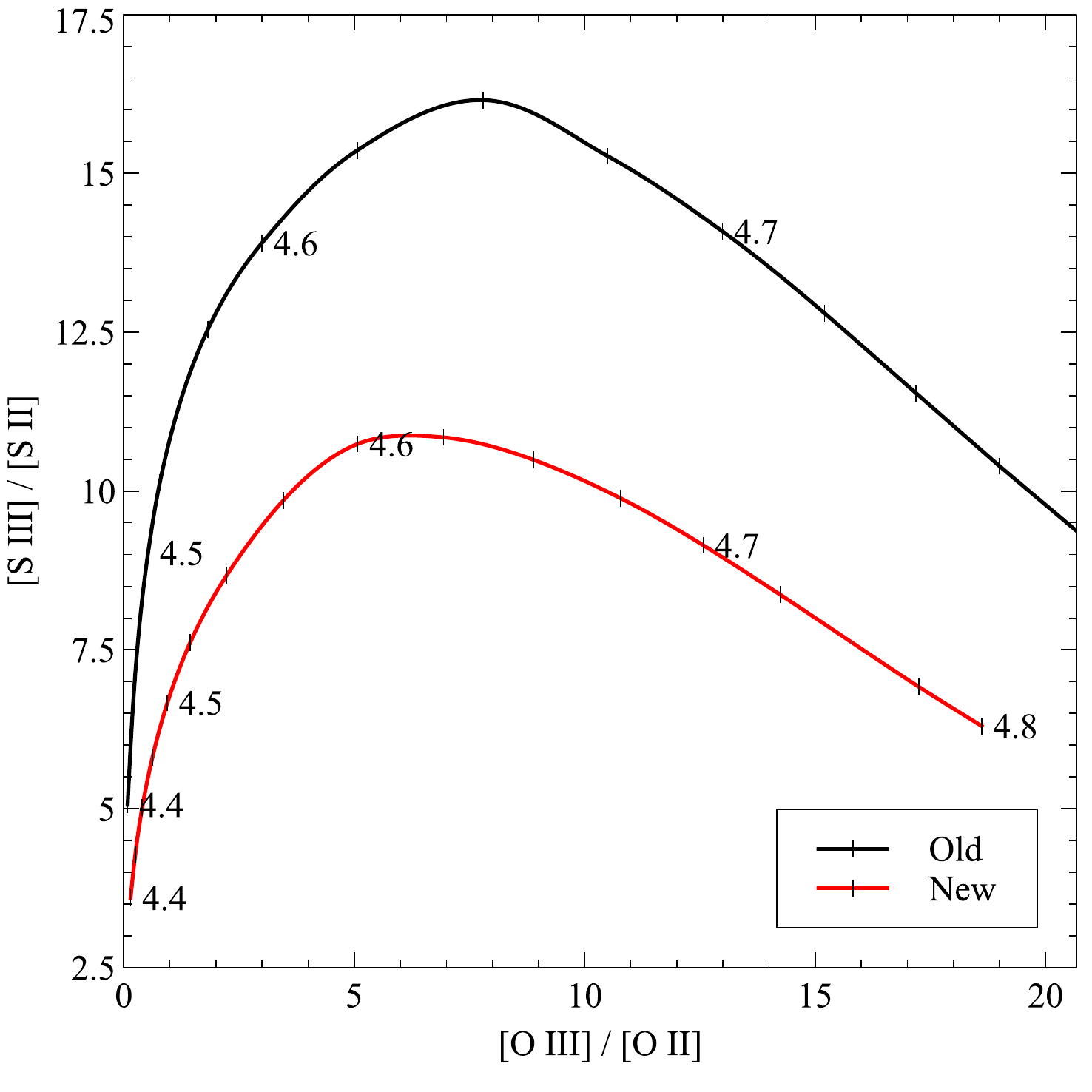}
\caption{
This shows the predicted ratios  
[S~III]  $\lambda$9530.6\AA /[S~II] $\lambda\lambda$ 6720\AA\ and
[O~III]  $\lambda$5006.84\AA /[O~II] $\lambda\lambda$ 3727\AA\
where [S~II] and [O~II] are the summed intensities of the doublet.
The model is of a blister H~II region with Orion abundances and dust,
ionized by various blackbodies with an ionization parameter of 
$\log U = -1.5$ \citep{2006agna.book.....O}.
The blackbody temperatures were varied and the log of the value 
is indicated on the curves.
The largest effect of the new recombination rate is to decrease
[S~III]/[S~II] at a given [O~III]/[O~II].
}
\label{figvaryT}
\end{figure}

The recombination rates recommended above have been adopted
in the development version of Cloudy \citep{2013RMxAA..49..137F} and
will be used in the next release, now scheduled for early 2015.
That version uses the general formula for DR suppression given by
\citet{Nikolic.D13Suppression-of-Dielectronic-Recombination-due-to-Finite}.
As recommended here, we do not suppress S$^{2+} \rightarrow$S$^+$ DR at
photoionized plasma temperatures.
These updates have a moderate effect on the spectrum.
Cloudy includes an extensive suite of test cases that are used to 
validate its performance.  
This includes Active Galactic Nuclei, planetary nebulae, H~II regions,
and other classes of object.
These tests show that the largest changes
occur for H~II regions ionized by cooler stars. 
The larger recombination rates increase the predicted [S~II] / [S~III],
ratio by as much as 50\% for late-type O stars.
Figure  \ref{figvaryT} shows typical results.
Changes in the  [O~II] and [O~III] lines
are modest and are caused by changes in the kinetic temperature
resulting from changes in the S spectra.  So the net effect is
that, at a given  [O~II]~/~[O~III] ratio, the [S~II]~/~[S~III] ratio is larger
by $\sim 20 - 50$\%.

This work was originally motivated by 
\citet{Henry.R12The-Curious-Conundrum-Regarding-Sulfur}, who documented
``a Curious Conundrum Regarding Sulfur Abundances in Planetary Nebulae''.
Uncertainties in the atomic database are a likely contributor to the conundrum.
This pointed us towards the S$^{2+} \rightarrow $S$^+$ recombination rate,
the only ion lacking modern DR calculations (e.g. missing from
http://amdpp.phys.strath.ac.uk/tamoc/DR/).
The revised DR rate derived here does not resolve the conundrum.
We can only speculate as to its origin, but the ion state of S in an H II region,
mainly single and doubly ionized, is lower than in a planetary nebula,
with its hotter central star ionizing S to higher levels. 
The stellar SED was a major concern in the present study, and it seems
likely that the SED for a planetary nebula nucleus might play a significant
role in resolving the conundrum.

\section{Conclusion}

This paper suggests a way to make progress in solving the vexing and long-standing problem of
finding good DR rates for the range of ions seen in astronomical observations.
An observational program, combined with spectral modeling and a parallel effort in
atomic theory, could make real progress in deriving DR rates for
third and fourth row elements with well defined uncertainties.
The elements of this program include the following:

\emph{Bootstrapping from astronomical observations of second-row elements}.  
Section 2  outlined a novel method in which spectral models are
used to match spectra produced by second-row elements, with good atomic data,
to infer the rate for an ion on the third row.
The key is that observations detect second-row ions which have a wider range
of ionization potential than the species with the unknown rate.
Section 3 shows that atomic theory can accommodate the empirical rate,
and we illustrated the uncertainty which remains as we move away from the
matching temperature.
We concluded with a fit that can be used over all temperatures of physical interest.
This procedure could be carried out for other species with adequate observations.

\emph{Insights from density dependencies}.  
Astronomical observations of objects covering a range of densities 
can provide additional insights to the atomic structure.
In the recombination process considered here the parent ion, S$^{2+}$, has a $^3$P ground term.
Our final fit to the DR rate predicts similar rate coefficients for the three $j=0,1,2$ levels
within the ground term (Table \ref{tableradrates}), but
this is not the case for results from other assumed atomic structures.
Figure \ref{figcompare} shows that some predict a strong dependency on $j$.
In those cases the total recombination rate will depend on the population of each $j$ level,
which in turn depends on the electron density.
So the recombination rate would have a strong density dependence if that structure was the correct one.
Observations of nebulae that cover a range of densities could check whether this
is the case, and could, absent the observations needed for the previous test,
decide which curve in Figure \ref{figcompare} matches the observations.

\acknowledgments
GJF acknowledges support by NSF (1108928, 1109061, and 1412155), NASA (10-ATP10-0053, 
10-ADAP10-0073, NNX12AH73G, and ATP13-0153), STScI (HST-AR- 13245, 
GO-12560, HST-GO-12309, GO-13310.002-A, and HST-AR-13914) and
is grateful to the Leverhulme Trust for support via the award of a 
Visiting Professorship at The Queen's University of Belfast (VP1-2012-025). 
TWG was supported in part by the NASA APRA grant NNX11AF32G.
NRB was supported in part the STFC UK APAP Network grant ST/J000892/1.

\clearpage

\bibliography{LocalBibliography}

\end{document}